\documentclass[12pt,preprint]{aastex}

\shorttitle{Filament Oscillations} \shortauthors{Li, et al.}

\begin{document}

%\title{Induced and Reflected Coronal EUV Waves on the Sun Observed with \emph{SDO}/AIA}
\title{\emph{SDO}/AIA Observations of Large-Amplitude Longitudinal Oscillations in a Solar Filament}

\author{Ting Li\altaffilmark{1} and Jun Zhang\altaffilmark{1}}

\altaffiltext{1}{Key Laboratory of Solar Activity, National
Astronomical Observatories, Chinese Academy of Sciences, Beijing
100012, China; [liting;zjun]@nao.cas.cn}

\begin{abstract}

We present the first \emph{Solar Dynamics Observatory}/Atmospheric
Imaging Assembly observations of the large-amplitude longitudinal
(LAL) oscillations in the south and north parts (SP and NP) of a
solar filament on 2012 April 7. Both oscillations are triggered by
flare activities close to the filament. The period varies with
filamentary threads, ranging from 44 to 67 min. The oscillations of
different threads are out of phase, and their velocity amplitudes
vary from 30 to 60 km s$^{-1}$, with a maximum displacement of about
25 Mm. The oscillations of the SP repeat for about 4 cycles without
any significant damping and then a nearby C2.4 flare causes the
transition from the LAL oscillations of the filament to its later
eruption. The filament eruption is also associated with a coronal
mass ejection and a B6.8 flare. However, the oscillations of the NP
damp with time and die out at last. Our observations show that the
activated part of the SP repeatedly shows a helical motion. This
indicates that the magnetic structure of the filament is possibly
modified during this process. We suggest that the restoring force is
the coupling of the magnetic tension and gravity.

\end{abstract}


\section{Introduction}

Oscillations of filaments have been observed using H$\alpha$
spectrograms since 1930s (Dyson 1930). They are broadly classified
into two groups according to their velocity amplitudes, namely, the
``large-amplitude oscillations" and ``small-amplitude oscillations"
(Thompson \& Schmieder 1991; Oliver \& Ballester 2002). The
large-amplitude oscillations occur when the entire filament
oscillates with a velocity amplitude of the order of 20 km s$^{-1}$
or larger. The small-amplitude oscillations are localised to part of
the filament, and have a velocity amplitude of the order of 2--3 km
s$^{-1}$ or even less. There have been many studies about
small-amplitude oscillations in the last few decades (Oliver 2009;
Ning et al. 2009; Soler et al. 2011). However, the large-amplitude
oscillations are less widely reported (Okamoto et al. 2004; Tripathi
et al. 2009; Arregui et al. 2012).

Moreover, most reported large-amplitude oscillations are
perpendicular to the filament axis. This kind of transverse
oscillations is often triggered by fast-mode magnetohydrodynamic
(MHD) waves coming from distant flares (Ramsey \& Smith 1966; Isobe
\& Tripathi 2006; Gilbert et al. 2008; Chen et al. 2008; Hershaw et
al. 2011; Liu et al. 2012). The restoring force in this type of
oscillations is generally thought to be the magnetic tension.

Unlike transverse oscillations, Jing et al. (2003) first reported
periodic motions along a solar filament axis by using high-cadence
H$\alpha$ observations. The longitudinal oscillations were triggered
by a subflare occurring near the filament footpoint. They considered
the restoring force due to magnetic tension. Afterwards, Jing et al.
(2006) and Vr\v{s}nak et al. (2007) observed similar large-amplitude
longitudinal (LAL) oscillations. Vr\v{s}nak et al. (2007) proposed
that the LAL oscillations were triggered by an injection of poloidal
magnetic field into the flux rope, and the restoring force was
caused by the magnetic pressure gradient along the filament axis.
Recently, Luna \& Karpen (2012) first simulated the LAL oscillations
and suggested that the restoring force was the projected gravity in
the flux tube dips where the threads oscillate. The oscillations
were strongly damped by the mass accretion of the threads. Zhang et
al. (2012) analyzed the LAL oscillations above the solar limb by
using multi-wavelength data. Their study suggested that the
oscillations were excited due to plasma injection and the restoring
force may be the gravity.

Since there are different viewpoints on the physical nature of LAL
oscillations, further investigations are needed. In this Letter, we
present the first observations of the LAL oscillations by the new
Atmospheric Imaging Assembly (AIA; Lemen et al. 2012) onboard the
\emph{Solar Dynamics Observatory} (\emph{SDO}; Pesnell et al. 2012).

\section{Observations and Data Analysis}

On 2012 April 7, \emph{SDO}/AIA observed LAL oscillations of an
active region (AR) filament in NOAA AR 11451 (N17W45). The LAL
oscillations occurred in the south and north parts (SP and NP in
Figure 1) of the filament (see Animations 1 and 3 in the online
journal). Ultimately, the SP of the filament erupted and the
eruption was associated with a B6.8 flare and a coronal mass
ejection (CME). The oscillations of the NP damped obviously with
time and died out at last.

The \emph{SDO}/AIA takes full-disk images in 10 (E)UV channels at
1.5$\arcsec$ resolution and high cadence of 12 s. Among the 10
wavelengths of AIA, the 171 and 304 {\AA} channels best show the LAL
oscillations and we focus on these channels in this study. The 171
{\AA} bandpass channel is dominated by emission of Fe IX formed at
$\sim$ 0.6 MK, and the 304 {\AA} channel dominated by He II emission
at 0.05 MK (O'Dwyer et al. 2010; Boerner et al. 2012; Parenti et al.
2012). The LAL oscillations were also observed by the Extreme
Ultraviolet Imager (EUVI; see Wuelser et al. 2004) aboard
\emph{Solar-Terrestrial Relations Observatory} (\emph{STEREO};
Kaiser et al. 2008) that was 111$\degr$ ahead of the Earth. EUVI
observes the chromosphere and corona in four spectral channels (304
{\AA}, 171 {\AA}, 195 {\AA} and 284 {\AA}) out to 1.7 $R_{sun}$ with
a pixel size of 1.6$\arcsec$. All the data used here are corrected
for solar rotation.

\section{Results}

\subsection{LAL Oscillations of the SP}

The SP of the analyzed filament mainly consists of two parts (Parts
1 and 2 in Figure 2\emph{a}), which have different axis directions
(slices A-B and C-D in Figure 2\emph{b}). By examining the
\emph{STEREO} A data, we find that Part 2 lies above Part 1 and
extends longer than Part 1.

At 13:26 UT, a weak flare-like brightening appeared in the vicinity
of the SP at STEREO A 304 {\AA} (see the white circle in Figure
1\emph{a}). The observations indicate that the subflare caused the
the activation of Part 1. The north end of Part 1 started to rise
slowly with a velocity of $\sim$ 20 km s$^{-1}$ since 13:27 UT and
the rise lasted about 5 min (Figure 2\emph{a}). After 13:32 UT, the
filament shrank southward along the axis of Part 1 (Figure
2\emph{b}). To illustrate the motion of filament, Figure 3 shows the
intensity evolution along slices A-B and C-D in Figure 2\emph{b}
(stack plots). Seen from the stack plot along slice A-B, the mass
motion of Part 1 has a velocity of 70 $\pm$ 10 km s$^{-1}$ (blue
dotted lines in Figure 3\emph{a}). Furthermore, the main body of
Part 1 shows a helical motion in the image plane during the interval
of the rise and shrinkage (refer to images between 13:27--13:34 UT
and Animations 2 and 4 in the online journal).

Following the mass motion of the filament, the oscillations along
Part 1 were clearly observed since 13:45 UT (Figure 3\emph{a}). At
about 14:06 UT, Part 2 started to oscillate along its axis (Figure
3\emph{b}). The arrows in Figure 3 indicate the times at which the
direction of the oscillating motions seem to reverse. It can be seen
that the oscillatory period varies with filamentary threads. Two
oscillatory profiles along slice A-B have oscillatory periods of
$\sim$ 44 and 54 min, respectively. Their oscillatory amplitudes of
the displacement obtained are $\sim$ 2.5 $\times$ 10$^{4}$ and 2.1
$\times$ 10$^{4}$ km, and the corresponding velocity amplitudes are
60 and 40 km s$^{-1}$, respectively. The LAL oscillation along slice
C-D shows a longer period of $\sim$ 67 min than that of Part 1. Its
maximum displacement of the oscillation is $\sim$ 2.0 $\times$
10$^{4}$ km, and the displacement shows no significant damping. The
oscillation in Figure 3\emph{b} has a velocity amplitude of about 30
km s$^{-1}$, and completes two cycles before it could be traced
clearly.

By comparing the three oscillatory profiles in Figure 3, we find
that the oscillations of different filamentary threads start
approximately in phase. However, due to different initiation times
and oscillatory periods, their phases are not synchronous. At 16:23
UT, partial filament material of Part 1 separated from the main body
and moved toward the south. The velocity of the mass motion is the
same as that at 13:32 UT (Figure 3\emph{a}). Meanwhile, similar
helical motion phenomenon was observed until 16:43 UT (see
Animations 2 and 4 in the online journal).

At the late phase of LAL oscillations, bi-directional flows along
the filament axis are observed. In order to display the flows
clearly, we select two parallel slices with a separation of 6.1 Mm
(slices E-F and G-H in Figure 2\emph{c}) and obtain their
corresponding stack plots in Figure 4. Between 16:34--16:50 UT, the
filament material located at slice E-F moved toward the north with a
velocity of $\sim$ 30 km s$^{-1}$, while some material at slice G-H
moved toward the south with velocities of 20--40 km s$^{-1}$ (the
region between two vertical dashed lines in Figure 4\emph{a}). Along
slice G-H, the flow toward the north was also observed, and the
velocity was about 50 km s$^{-1}$. Between 17:12--17:36 UT, the
oscillating material moved southwardly along slice E-F, and its
velocity was about 20 km s$^{-1}$. While the material at slice G-H
moved northwardly with a velocity of $\sim$ 10 km s$^{-1}$ (the
region between two blue dashed lines in Figure 4\emph{a}). The
oscillations of the SP last about 4 hr, almost 4 times the
corresponding periods, before the onset of the filament eruption.
%``A"
\subsection{Eruption of the SP}

At about 16:59 UT, a C2.4 flare occurred at the northeast of the SP
(Figure 1\emph{b}), which peaked at 17:08 UT and ended at 17:15 UT.
The two-ribbon flare was associated with no filament eruption and
CME. The flare resulted in the southward mass motion along the axis
of the SP (see Animations 1 and 3 in the online journal). About 30
min after the flare, the oscillating filament started to erupt
toward the southwest in AIA images and the eruption had a velocity
range from 60 km s$^{-1}$ to 120 km s$^{-1}$ (Figure 4). As seen in
the inner (COR1) and outer (COR2) coronagraphs onboard the
\emph{STEREO} A, the filament eruption finally resulted in a CME
with an initial velocity of 200 km s$^{-1}$ in the COR1's filed of
view (FOV). The SP eruption was also associated with a B6.8 flare,
which started at 19:08 UT, peaked at 19:55 UT and ended at 20:16 UT.

During the eruptive phase, several signatures that may suggest
magnetic reconnection between filament threads were observed.
Partial filament material moved in the opposite direction (from N to
M in Figure 2\emph{c}) while the entire filament started to erupt
southwardly at 17:45 UT. Meanwhile, the region that the material
flowed into displayed faint localized brightenings. To derive the
inflow speed of the material, we used the stack plot shown in Figure
4\emph{c}. Multi-threads that moved along slice M-N could be
identified and the movements of the treads are recognized as bright
lines in the stack plot (marked with black dashed lines). The
apparent velocity of the inflow is about 120 km s$^{-1}$. At about
18:10 UT, intense and concentrated brightenings started near the
position that the filament threads flowed into previously (Figure
2\emph{d}).

\subsection{LAL Oscillations of the NP}

About 2 min after the C2.4 flare occurring at the middle part of the
entire filament (17:02 UT), the brightenings of filamentary threads
nearby the flare ribbons were clearly observed (see Animations 1 and
3 in the online journal). Subsequently, the north filament became
activated and the northward mass motion along the filament axis was
formed since 17:14 UT. From $\sim$ 17:20 UT, the great bulk of
filament material coming from the south was injected to the NP of
the filament (Figure 5a). Initially, the injected material moved
toward the east and the average velocity was about 60 km s$^{-1}$.
Then it changed its direction and turned back to the west upon
reaching its maximum displacement at $\sim$ 18:24 UT (Figure
5\emph{c}).

The movement of the oscillating material is nearly along the axis of
the NP, which is also the LAL oscillation. The oscillations have a
period of $\sim$ 57 min and last about 3 hr, almost three cycles
before it damps out at last.

\section{Summary and Discussion}

In this Letter, we report the first simultaneous observations of the
LAL oscillations of the filament and its later eruption. About 30
min before the filament eruption, a C2.4 flare occurred at the north
of the SP. The flare resulted in the southward mass motion along the
axis of the SP. This may suggest that the flare causes the
transition from oscillations to the eruption. The flare may impose a
strong impulse on the oscillating filament and leads to its loss of
equilibrium that results in the filament eruption. The filament
eruption was also associated with a CME and a B6.8 flare. Chen et
al. (2008) and Bocchialini et al. (2011) also presented observations
of vertical and transverse oscillations prior to eruption. They
suggested that filament oscillations could be considered as another
precursor of CMEs.

The LAL oscillations occur in the SP and NP of the filament, and
they have some common characteristics. They are both triggered by
flare activities close to the filament (Jing et al. 2003; Vr\v{s}nak
et al. 2007). These flares may excite the mass motion along the
filament axis and then the longitudinal oscillations are formed. The
oscillations of different threads are out of phase, and their
velocity amplitudes vary from 30 to 60 km s$^{-1}$, with a maximum
displacement of about 25 Mm. The oscillations of the SP repeat for
about 4 cycles without any significant damping before the onset of
the filament eruption. For the NP, the oscillations damp with time
and die out at last.

Generally, there are several physical mechanisms that have been
proposed to explain various oscillations. Now we compare these wave
modes (such as magnetic kink-mode wave, Alfv\'{e}nic oscillations
and longitudinal slow magentosonic waves) with our observational
results. The transverse oscillations of filaments and coronal loops
are interpreted as standing fundamental kink modes (Aschwanden et
al. 1999; Ofman \& Wang 2008; Hershaw et al. 2011), and the
oscillating amplitude of the location close to the filament/loop
footpoints is smaller than that of other locations. However, the
variation of the amplitude with locations is not observed in our
event (Figure 3a) and the kink-mode oscillations can be ruled out.
The direction of Alfv\'{e}nic oscillations is perpendicular to the
direction of the magnetic field. However, the LAL oscillations of
the filaments are approximately along the direction of magnetic
field. Thus the LAL oscillations should not be Alfv\'{e}nic
oscillations. The propagating intensity disturbances in polar plumes
and coronal loops are generally interpreted as slow magentosonic
waves (Ofman et al. 1999, 2000, 2012; De Moortel et al. 2000; Wang
2011). In our event, we find no significant phase propagation along
the filament and thus the longitudinal slow magentosonic waves are
also ruled out.

We suggest that the restoring force is the coupling of the magnetic
tension and gravity. Our observations show that the activated part
of the SP repeatedly shows a helical motion. This indicates that the
magnetic structure of the filament is possibly modified during this
process. The twisted field of the filament has the component
perpendicular to the direction of the filament axis although it is
mostly parallel to the filament axis. The motions of filamentary
threads lead to distortion of the twisting magnetic field, and the
field component perpendicular to the direction of the filament axis
generates magnetic tension along the filament axis as the restoring
force, similar to the analysis of Jing et al. (2003). The gravity
could also be considered as the restoring force. Recently, Luna \&
Karpen (2012) and Zhang et al. (2012) simulated LAL oscillations and
suggested that the restoring force was the projected gravity in the
flux tube dips where the threads oscillate. The comprehensive
understanding of the physical nature on the LAL oscillations of
filaments needs more observations and further analysis in the
future. \textbf{Assuming the restoring force is only the gravity, we
estimate the minimum magnetic field strength in the filament as
28--55 G according to the analysis of Luna \& Karpen (2012). If the
restoring force is only the magnetic tension, the estimated magnetic
field strength is about 32 G based on the study of Kleczek \&
Kuperus (1969). These estimated magnetic field strengths of the
filament are both similar to the observations of active region
filament magnetic fields (Mackay et al. 2010), so it is difficult to
determine which kind of force is dominant as the restoring force. We
consider that there exists great uncertainty if we employ LAL
oscillations of filaments for corona seismology.}

We find that the oscillatory period varies with filamentary threads,
ranging from 44 to 67 min. This discrepancy may suggest that the
magnetic field and plasma condition of filamentary threads are
different (Hershaw et al. 2011). Due to different initiation times
and oscillatory periods, the oscillatory phases of different
filamentary threads are not synchronous, which indicates that the
excited threads lose their coherence from the beginning.

Bi-directional flows along the filament axis are observed at the
late phase of LAL oscillations with a speed range of 10--40 km
s$^{-1}$ (Figures 4\emph{a} and \emph{b}). These speeds are a factor
of 2 greater than those of the counter-streaming flows reported
before (5--20 km s$^{-1}$; Zirker et al. 1998; Schmieder et al.
2010; Lin 2011). Since the oscillations of different filamentary
threads are not synchronous, the interaction between filamentary
threads is ultimately destroyed at the late phase. Thus some threads
have opposite movement directions, which shows the characteristic of
bi-directional flows. Our observations show that the bi-directional
flows occur before the filament eruption, similar to the
observations of Lin et al. (2003) and Schmieder et al. (2008).

\acknowledgments {We are grateful to Dr. P. F. Chen for useful
discussions. We acknowledge the SECCHI and AIA for providing data.
This work is supported by the National Basic Research Program of
China under grant 2011CB811403, the National Natural Science
Foundations of China (11025315, 40890161, 10921303 and 11003026),
the CAS Project KJCX2-EW-T07, and the Young Researcher Grant of
National Astronomical Observatories, Chinese Academy of Sciences.}

{}
\clearpage

\begin{figure}
\centering
\includegraphics
[bb=72 181 500 644,clip,angle=0,scale=0.9]{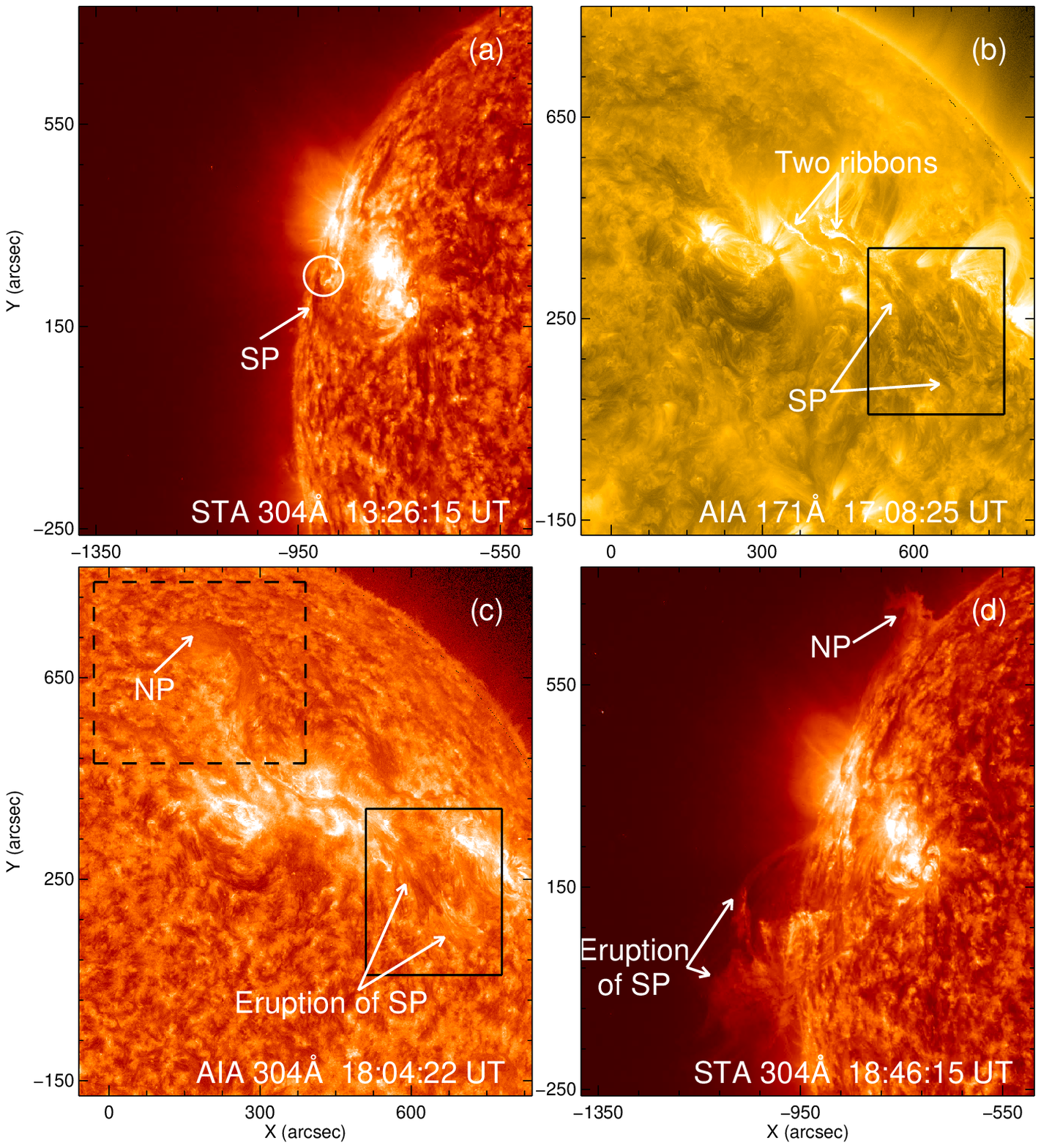} \caption{
{\emph{SDO}}$/$AIA 171 {\AA}, 304 {\AA} and {\emph{STEREO}} A 304
{\AA} images showing the evolution of the oscillating filament
(including the SP and NP) on 2012 April 7 (see Animations 1 and 3,
available in the online edition of the journal). The white circle in
panel \emph{a} highlights the subflare before the onset of
oscillations. The solid rectangles denote the FOV of Figure 2 and
the dashed rectangles denote the FOV of Figure 5. \label{fig1}}
\end{figure}
\clearpage

\begin{figure}
\centering
\includegraphics
[bb=45 130 520 690,clip,angle=0,scale=0.75]{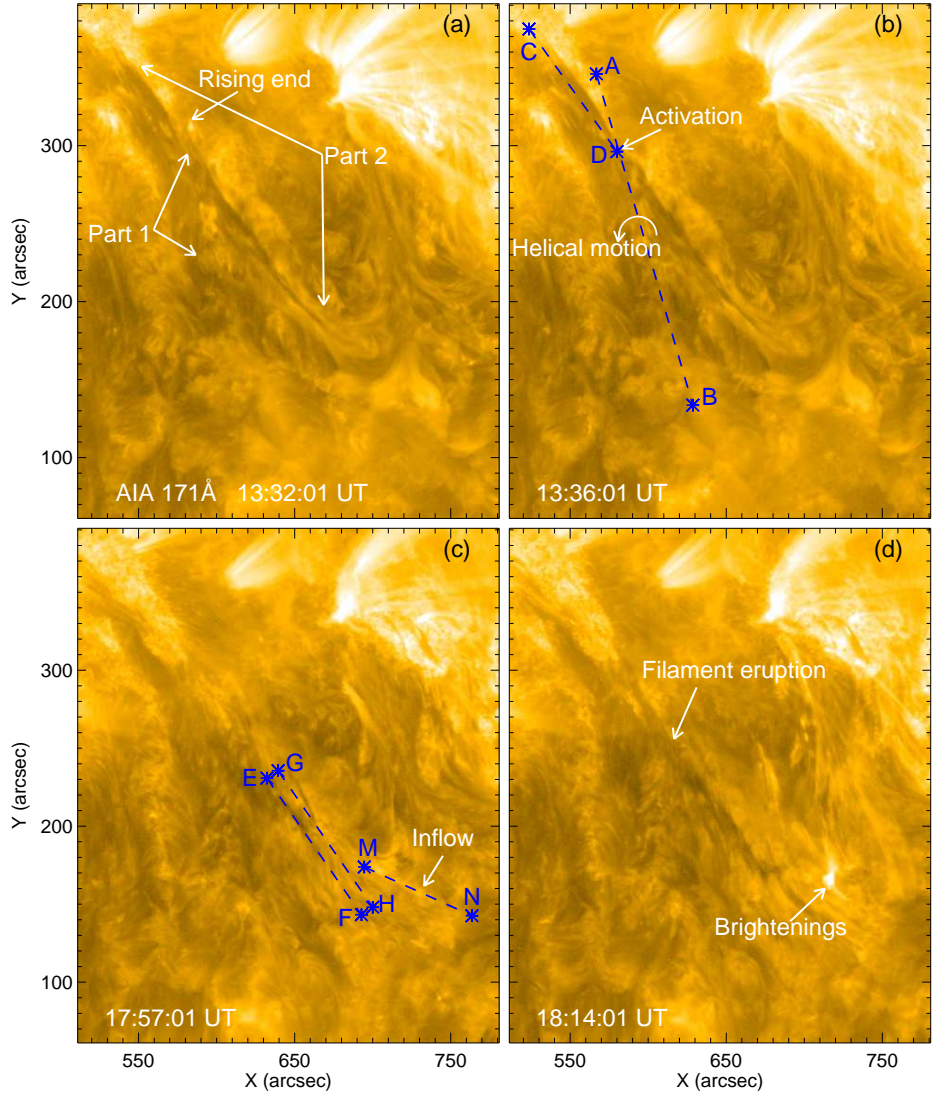}
\caption{{\emph{SDO}}$/$AIA 171 {\AA} images showing the early and
late evolution of the SP (see Animations 2 and 4 available in the
online edition of the journal). Blue dashed slices A-B and C-D in
panel \emph{b} are used to obtain the stack plots shown in Figure 3.
Blue dashed slices E-F, G-H and M-N in panel \emph{c} are used to
obtain the stack plots shown in Figure 4. \label{fig3}}
\end{figure}
\clearpage

\begin{figure}
\centering
\includegraphics
[bb=108 271 470 557,clip,angle=0,scale=1.2]{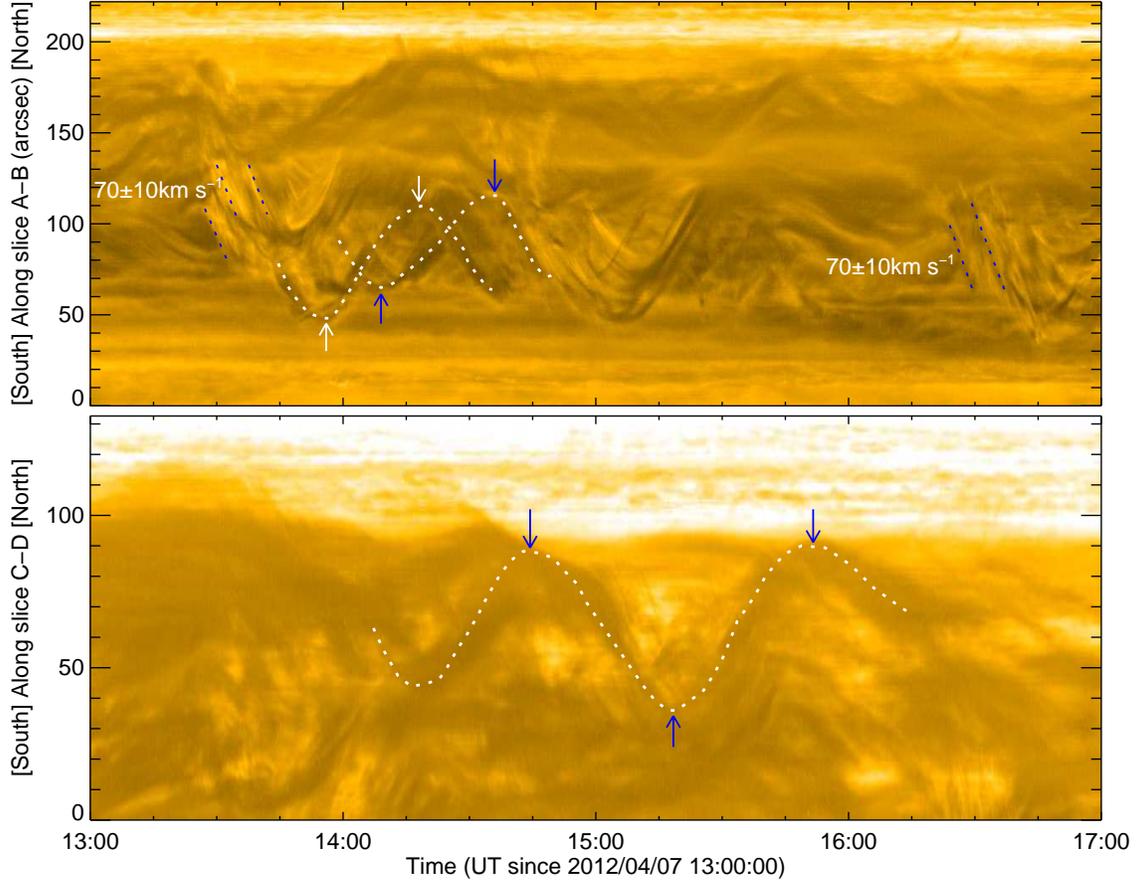} \caption{171
{\AA} original stack plots along slices A-B and C-D indicated in
Figure 2\emph{b}. Blue dotted lines in panel \emph{a} denote the
shrinkage of filament material. White dotted curves represent
oscillatory profiles along slices A-B and C-D. The arrows point to
the reversal points of the oscillation. \label{fig1}}
\end{figure}
\clearpage

\begin{figure}
\centering
\includegraphics
[bb=113 197 470 616,clip,angle=0,scale=1.0]{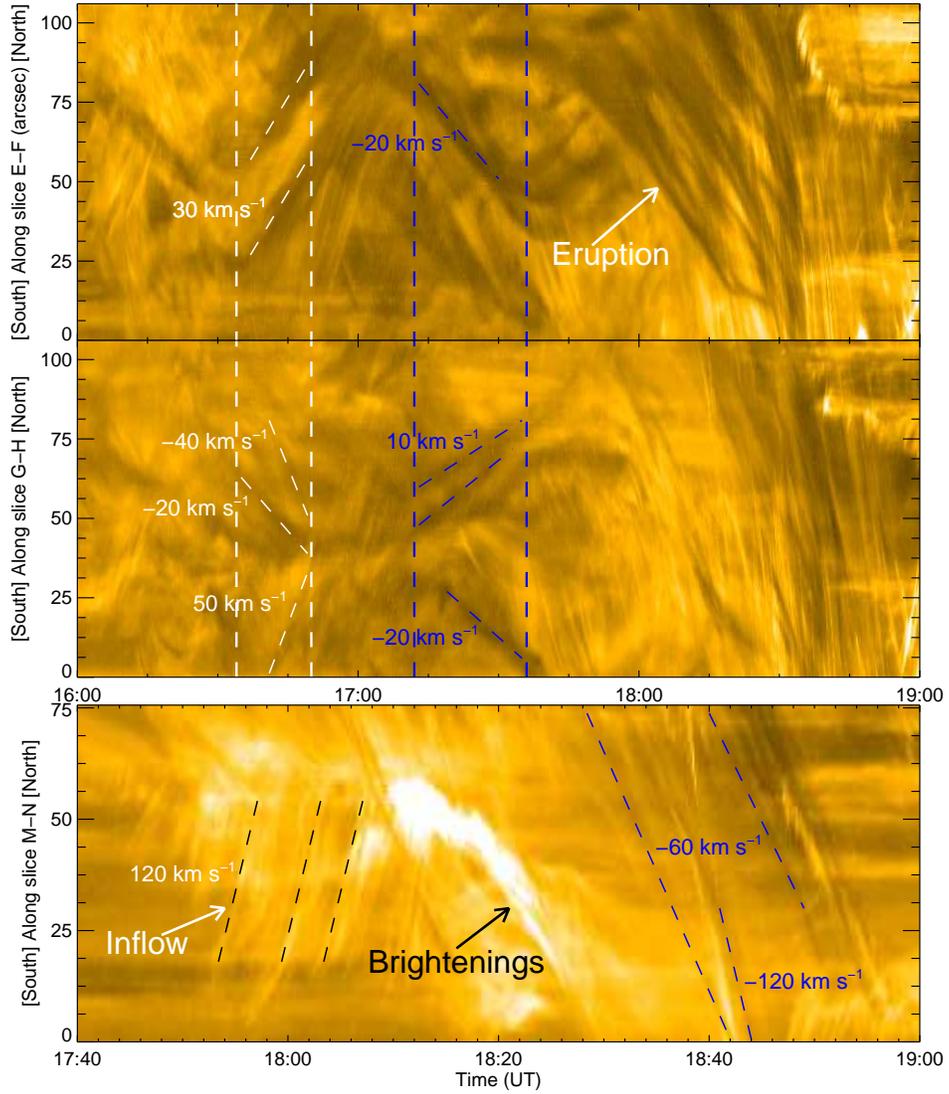} \caption{171
{\AA} original stack plots along slices E-F and G-H (panels \emph{a}
and \emph{b}) showing the bi-directional flows and the stack plot
along slice M-N (panel \emph{c}) showing the inflows toward the
north. The positions of the slices are indicated in Figure
2\emph{c}. \label{fig4}}
\end{figure}
\clearpage

\begin{figure}
\centering
\includegraphics
[bb=94 278 476 579,clip,angle=0,scale=1.1]{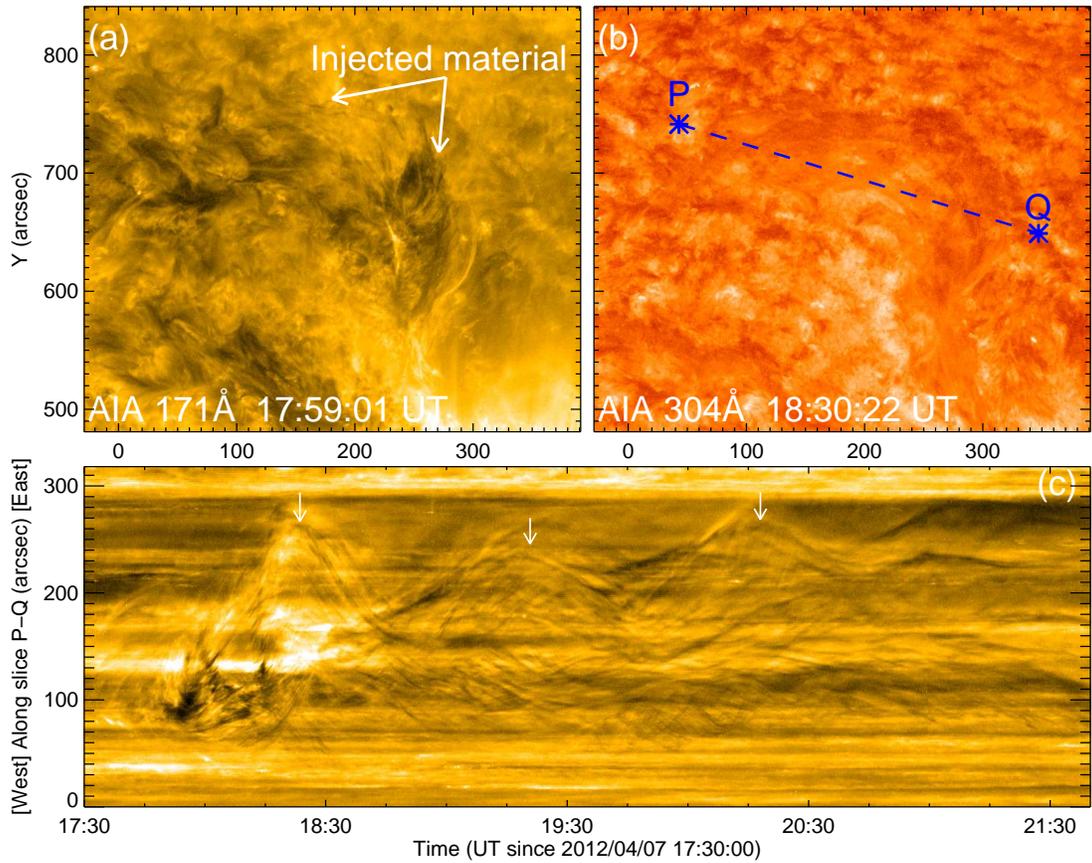}
\caption{{\emph{SDO}}$/$AIA 171 {\AA} (panel \emph{a}) and 304 {\AA}
(panel \emph{b}) images showing the NP of the filament, and 171
{\AA} original stack plot (panel \emph{c}) along slice P-Q. The
arrows in panel \emph{c} denote the reversal points of the
oscillation. \label{fig4}}
\end{figure}
\clearpage

\end{document}